\documentclass[twocolumn]{article}
\usepackage[utf8]{inputenc} % allow utf-8 input
\usepackage[T1]{fontenc}    % use 8-bit T1 fonts

\usepackage{amsmath,amsfonts}
\usepackage{algorithmic}
\usepackage{algorithm}
\usepackage{array}
\usepackage[caption=false,font=footnotesize]{subfig}
\usepackage{textcomp}
\usepackage{stfloats}
\usepackage{url}
\usepackage{verbatim}
\usepackage{graphicx}
\usepackage{siunitx}

\title{Embedded Model Control of Networked Control Systems: an Experimental Case-study - Stability analysis and further results}

\date{}

\author{Luca Nanu\thanks{Department of Mechanical and Aerospace Engineering, Polytechnic of Turin, Corso Duca degli Abruzzi 24, Turin, 10129, Italy. Mail address: \textit{luca.nanu@polito.it}} ,
Carlos Perez Montenegro\thanks{With Polytechnic of Turin, Corso Duca degli Abruzzi 24, Turin, 10129, Italy. Mail address: \textit{carlos.perez@polito.it}} ,
Luigi Colangelo\thanks{Department of Electronics and Telecommunications Engineering, Polytechnic of Turin, Corso Duca degli Abruzzi 24, Turin, 10129, Italy. Mail address: \textit{luigi.colangelo@polito.it}} ,
Carlo Novara\thanks{Department of Electronics and Telecommunications Engineering, Polytechnic of Turin, Corso Duca degli Abruzzi 24, Turin, 10129, Italy. Mail address: \textit{carlo.novara@polito.it}}
}

\begin{document}

\maketitle
	
\begin{abstract}
	In Networked Control Systems (NCS), the absence of physical communication links in the loop leads to relevant issues, such as measurement delays and asynchronous execution of the control commands. These issues may lead to unwanted control behaviours. This ArXiv paper is intended to give additional results to the work presented in \cite{Nanu2022}. The last one presents an original approach, based on the Embedded Model Control, to deal with experimental scenarios characterized by asynchronous control timing. The effectiveness of the proposed approach is demonstrated with a differential-drive robot, first with high-fidelity simulations and finally with several experimental tests.
	
	Specifically, the present work aims to study the stability analysis of the EMC experimental setup and to give further experimental results, to complement those presented in the main paper, \cite{Nanu2022}.
\end{abstract}
	
\section{Introduction}

At the beginning, the present paper provides additional studies on the stability analysis for the specific experimental test case with the differential-drive robot presented in \cite{Nanu2022}, with the introduction of the asynchronous control timing.

Furthermore, additional results are studied in the final part of the present paper, comprising:

(a) an experimental test comparing the estimated output and the measured output for the DC motor speed. In this test, it is evident how the model error is very small, thus verifying the effectiveness of the EMC observer state estimation;

(b) a set of experimental tests in which the EMC control unit is tested in several critical conditions concerning the asynchronous timing, in order to explore the control space and to highlight the boundaries of the control action;

(c) further benchmarking tests, between EMC and a Proportional-Integrative (PI) control.
	
\section{Mobile robot application: EMC Stability analysis}

This section gives a stability proof of the EMC control unit applied to the experimental differential-drive robot setup. In addition, the stability analysis of the EMC unit in a general control scenario is already treated in \cite{novara2016control}.

The EMC unit is split into sub-models and a stability analysis is made for each of them. The idea is that the EMC unit can guarantee internal and bounded-input-bounded-output (BIBO) stability of the closed-loop system if all the sub-models are internal and BIBO stable.

	\subsection{Internal model stability proof}

The internal model of the EMC unit can be formulated in the form of state equations as follows, \cite{Canuto2007}

\begin{equation}
	\label{eq:DTmdl} 
	\begin{aligned}
		&\mathbf{x}(k+1) = \begin{bmatrix} \mathbf{x}_c \\ \mathbf{x}_d \end{bmatrix}(k+1) =
		\underbrace{\begin{bmatrix} \mathbf{A}_c & \mathbf{H}_c \\ 0 & \mathbf{A}_d \end{bmatrix}}_\mathbf{A}
		\mathbf{x}(k) \ + \\
		&+ \underbrace{\begin{bmatrix} \mathbf{B}_c \\ \mathbf{B}_d \end{bmatrix}}_\mathbf{B} \mathbf{u}(k) + \underbrace{\begin{bmatrix} \mathbf{G}_c \\ \mathbf{G}_d \end{bmatrix}}_\mathbf{G} \mathbf{\overline{w}}(k), \\
		&\mathbf{y}_m(k) = \underbrace{\begin{bmatrix} \mathbf{C}_c \ \mathbf{C}_d \end{bmatrix}}_\mathbf{C} \mathbf{x}(k),
	\end{aligned}
\end{equation}

where $\mathbf{x}_c$ and $\mathbf{x}_d$ refer to the controllable (or canonical) and rejector states, respectively.

On the other side, the matrices $ \mathbf{A}, \mathbf{B}, \mathbf{C} $, are related respectively to: (i) the state $ \mathbf{x} $, (ii) the control input $ \mathbf{u} $, and (iii) the estimated output from the internal model $ \mathbf{y}_m $. In addition the matrix $ \mathbf{G} $ is present, for the estimation of the noise vector $ \mathbf{\overline{w}} $. In particular, the matrix $ \mathbf{A} $ is made by the sub-matrices: $ \mathbf{A}_c $, related to $ \mathbf{x}_c $, $ \mathbf{A}_d $ related to $ \mathbf{x}_d $, and $ \mathbf{H}_c $, which considers the interactions of $ \mathbf{x}_d $ on $ \mathbf{x}_c $. Similarly, we have the sub-matrices $ \mathbf{B}_c$ and $\mathbf{B}_d $ for $ \mathbf{B} $ matrix, $ \mathbf{G}_c$ and $\mathbf{G}_d $ for $ \mathbf{G} $ matrix, $ \mathbf{C}_c $ and $ \mathbf{C}_d $ for $ \mathbf{C} $ matrix.

Equation \eqref{eq:DTmdl} for the internal model can be converted from time to frequency domain, as follows

\begin{equation}
	\label{eq:freq_int_mdl}
		\begin{aligned}
			\mathbf{X}(z) = & \underbrace{\{ z \mathbb{I} - \left[ \mathbf{A} - \mathbf{GLC} \right] \}^{-1}}_{\mathbf{X}_{zi}(z)} \mathbf{X}(0) + \\
			& + \underbrace{\mathbf{B} \{ z \mathbb{I} - \left[ \mathbf{A} - \mathbf{GLC} \right] \}^{-1}}_{\mathbf{X}_{zs,1}(z)} \mathbf{U}(z) + \\
			& + \underbrace{\mathbf{GL} \{ z \mathbb{I} - \left[ \mathbf{A} - \mathbf{GLC} \right] \}^{-1}}_{\mathbf{X}_{zs,2}(z)} \mathbf{Y}(z)
		\end{aligned}
\end{equation}
where $ \mathbf{X}_{zi}(z) $ denotes the zero-input response of the closed-loop system when $ \mathbf{U}(z) = \mathbf{Y}(z) = 0 $, $ \mathbf{X}_{zs,1}(z) $ the zero state response when $ \mathbf{X}(0) = \mathbf{Y}(z) = 0 $ and $ \mathbf{X}_{zs,2}(z) $ the zero state response when $ \mathbf{X}(0) = \mathbf{U}(z) = 0 $.

Both the internal and BIBO stability of the internal model can be studied by imposing $ \det \{ z \mathbb{I} - \left[ \mathbf{A} - \mathbf{GLC} \right] \}^{-1} $ equal to a characteristic polynomial $ (z - p_1) (z - p_2) \dots (z - p_n) $, where $ n $ is the $ \mathbf{A} $ matrix dimension: in order to guarantee the asymptotic stability, the poles $ \left[ p_1, p_2, \dots, p_n \right] $ must lie inside the unitary circle. From the equation above it is clear that the stability finally depends on the observer matrix $ \mathbf{L} $ components' design, which will be discussed in Section \ref{sec:noise_est_stab}.

	\subsection{Plant stability proof}

The plant of the DC motor, i.e. the mathematical description of the DC motor physical behaviour, must guarantee BIBO stability. Equation \eqref{eq:Second order disturbance_eq_DT} define the state equations for the DC motor

\begin{equation}
	\label{eq:Second order disturbance_eq_DT}
		\begin{aligned}
			& \begin{cases}
				& \omega(k+1) = \left( -\frac{1}{\tau_m}T_s+1 \right)  \omega(k) + \frac{T_s}{\tau_m k'_v} V_a(k) + d(k), \\
				& x_{d1}(k+1) = (T_s+1) x_{d1}(k) + T_s \left[ x_{d2}(k) + \overline{w}_2(t) \right], \\
				& x_{d2}(k+1) = (T_s + 1) x_{d2}(k) + T_s \overline{w}_3(k),
			\end{cases} \\
			& d(k)= T_s\left[ \overline{w}_1(k) + x_{d1}(k)\right], \\
			& \omega(0) = \omega_0, \quad x_{d1}(0) = x_{d10}, \quad x_{d2}(0) = x_{d20},
		\end{aligned}
\end{equation}

where $ \omega $ is the DC motor output speed, $ \tau_m $ is the DC motor mechanical time constant, $ k'_v $ is the back electromotive force constant specifically found for the studied motor, $ [\overline{w}_1, \overline{w}_2, \overline{w}_3] $ are the components of the noise vector $ \mathbf{\overline{w}} $, $ [x_{d1},x_{d2}] $ are the components of the disturbance state $ \mathbf{x}_d $, $ T_s $ is the sampling time.

Equation \eqref{eq:Second order disturbance_eq_DT} can be translated in frequency domain and reshaped in a matrix form, i.e.

\begin{equation}
	\label{eq:freq_mot_eqs}
	\begin{aligned}
		& W_{_M}(z) = \frac{Y(z)}{U(z)} = \frac{\beta_0 (z+1) }{z^2+\alpha_1 z + \alpha_0}, \\
		& \beta_0 = \frac{T^2}{k_v (\tau_m T + 2 \tau_m \tau_a)}, \\
		& \alpha_0 = \frac{T^2 - \tau_m T + 2 \tau_m \tau_a}{\tau_m T + 2 \tau_m \tau_a}, \\
		& \alpha_1 = \frac{T^2 - 4 \tau_m \tau_a}{ \tau_m T + 2 \tau_m \tau_a}
	\end{aligned}
\end{equation}

The worst case for the stability corresponds to the lowest sampling time possible. For example, by considering for $ T_s $ a range of $ [0.01, 0.03] \ \si{\second} $, the lowest sampling time corresponds to $ T_s = 0.01 \ s $: hence the parameters become $ \beta_0 = 0.1313, \alpha_0 = -0.1471, \alpha_1 = -0.4835 $.

The motor transfer function can be alternatively expressed as

\begin{equation}
	W_{_M}(z) = \frac{\beta_0 (z + 1)}{\left( z - d_0 \right) \left( z - d_1 \right)}
\end{equation}

where the two poles are $ d_0 = 0.695, d_1 = -0.212 $, both inside the unitary circle, leading to asymptotic BIBO stability of the motor plant transfer-function.

	\subsection{Reference dynamics stability proof}

The reference dynamics equations are recalled from \cite{Nanu2022}

\begin{equation}
	\label{eq:FB_refdyn}
	\begin{aligned}
		& \overline{\mathbf{x}}(k+1) = \mathbf{A}_{_R}\overline{\mathbf{x}}(k) +  \mathbf{B}_{_R} \overline{\mathbf{r}}(k), \\
		& \overline{\mathbf{y}}(k) =  \mathbf{C}_{_R}\mathbf{\overline{x}}(k), \\
		& \mathbf{A}_{_R}= \mathbf{A}_c - \mathbf{B}_c \mathbf{K}_{_R}, \\
		& \mathbf{B}_{_R} = \mathbf{B}_c \mathbf{N}_{_R}, \quad \mathbf{C}_{_R} = \mathbf{C}_c.
	\end{aligned}
\end{equation}

The reference trajectory $ \mathbf{\overline{r}}(k) $ is obtained by exploiting a static-state feedback, with as nominal control input $ \overline{\mathbf{u}}(k) = -\mathbf{K}_{_R} \overline{\mathbf{x}}(k) + \mathbf{N}_{_R} \overline{\mathbf{r}}(k) $, where $\mathbf{K}_{_R} = k_{_R}$ and $\mathbf{N}_{_R} = n_{_R}$ are the reference generator gains.

The reference dynamics internal and BIBO stability can be studied by exploiting \eqref{eq:FB_refdyn} into frequency domain:

\begin{equation}
	\label{eq:freq_ref_dyn}
	\mathbf{\overline{X}}(z) = \underbrace{z \left[ z \mathbb{I} - \mathbf{A}_{_R} \right]^{-1}}_{\mathbf{\overline{X}}_{zi}(z)} \mathbf{X}(0) + \underbrace{\mathbf{B}_{_R} \left[ z \mathbb{I} - \mathbf{A}_{_R} \right]^{-1}}_{\mathbf{\overline{X}}_{zs}(z)} \mathbf{\overline{R}}(z)
\end{equation}

where $ \mathbf{\overline{X}}_{zi}(z) $ denotes the zero-input response (hence when $ \mathbf{R}(z) = 0 $), $ \mathbf{\overline{X}}_{zs}(z) $ the zero state response when $ \mathbf{X}(0) = 0 $, $ \mathbf{\overline{R}} $ the reference trajectory input $ \mathbf{\overline{r}} $ translated in frequency domain. The stability depends on the term $ \left[ z \mathbb{I} - \mathbf{A}_{_R} \right]^{-1} $. Indeed, by applying the pole-placement technique, the continuous-time eigenvalue $ \mathbf{A}_{_R} = a_{_R} $ is user-defined to lie in the right-half plane, to guarantee stability in frequency-domain.

\subsection{Control law stability proof}

The controller block is part of an external closed-loop system, including a PI controller and the internal model controllable dynamics.

The EMC control law is made up by three terms: (i) the feed-forward component $\overline{\mathbf{u}}$, (ii) the state feedback $ \mathbf{u}_{trk} = \mathbf{K}_{_C}\overline{\mathbf{e}} $, and (iii) the disturbance rejection term $ \mathbf{u}_d = \mathbf{M}\mathbf{x}_d $. Ergo, being $\mathbf{\overline{e}}$ the tracking error, the full command expression holds:
\begin{equation}
	\label{eq:u_law}
	\begin{split}
		\mathbf{u}(k) &= \overline{\mathbf{u}}(k) + \mathbf{K}_{_C}\overline{\mathbf{e}}(k) - \mathbf{M}\mathbf{x}_d(k), \\
		\overline{\mathbf{e}}(k) &= \left( 	\overline{\mathbf{x}}(k) - \mathbf{Q} \mathbf{x}_d(k) \right) - \mathbf{x}_c(k).
	\end{split}
\end{equation}

The control block state equations are defined as follows, \cite{Nanu2022}

\begin{equation}
	\label{eq:CLsys}
	\begin{aligned}
		& \begin{bmatrix} \mathbf{x}_1 \\ \mathbf{x}_2 \end{bmatrix}(k+1) = \mathbf{A}_{ctrl}
		\begin{bmatrix} \mathbf{x}_1 \\ \mathbf{x}_2 \end{bmatrix}(k) + \mathbf{B}_{ctrl} \mathbf{\overline{e}}(k), \\
		& \mathbf{A}_{ctrl} = \begin{bmatrix} -\mathbf{A}_c - k_p \mathbf{B}_c & k_i \mathbf{B}_c \\ -1 & 1
		\end{bmatrix}, \\
		& \mathbf{B}_{ctrl} = \begin{bmatrix} k_p \mathbf{B}_c \\ 1 \end{bmatrix}, \\
		& \mathbf{A}_c = -\frac{T_s}{\tau_m} + 1, \ \ \mathbf{B}_c = \frac{T_s}{\tau_m k'_v}
	\end{aligned}
\end{equation}

where $ \mathbf{x}_1 = \mathbf{x}_c $ is the already defined controllable state of the internal model and $ \mathbf{x}_2 $ is the augmented state introduced by the PI controller.

Both the internal and BIBO stability can be studied following \eqref{eq:CLsys}, translated in frequency domain:

\begin{equation}
	\label{eq:freq_CLsys}
	\begin{aligned}
		\mathbf{X}_{ctrl}(z) = & \underbrace{z \left[ z \mathbb{I} - \mathbf{A}_{ctrl} \right]^{-1}}_{\mathbf{X}_{ctrl,zi}(z)} \mathbf{X}_{ctrl}(0) + \\
		& + \underbrace{\mathbf{B}_{ctrl} \left[ z \mathbb{I} - \mathbf{A}_{ctrl} \right]^{-1}}_{\mathbf{X}_{ctrl,zs}(z)} \mathbf{\overline{e}}(z)
	\end{aligned}
\end{equation}
where $ \mathbf{X}_{ctrl,zi}(z) $ denotes the zero-input response (hence when the tracking error $ \mathbf{\overline{e}}(z) = 0 $), $ \mathbf{X}_{ctrl,zs}(z) $ the zero state response when $ \mathbf{X}_{ctrl}(0) = 0 $. Similarly as before, the stability analysis depends on the term $ \left[ z \mathbb{I} - \mathbf{A}_{ctrl} \right]^{-1} $, which can be solved by applying the pole-placement technique.

The disturbance rejection input term $ \mathbf{u}_d = \mathbf{M} \mathbf{x}_d $ can be designed to guarantee the stability by finding a suitable matrix $ \mathbf{M} $. This matrix can be obtained by following the Sylvester-Francis equation, \cite{Canuto2007,Acuna-Bravo2017}.

	\subsection{Noise estimator stability proof}
	\label{sec:noise_est_stab}
The noise estimator block goal is to achieve the closed-loop stabilization of the predictor as well as to ensure a suitable disturbance estimation capability, i.e. estimate the noise vector $ \mathbf{\overline{w}} $. The noise estimator block governing equations are as follows

\begin{equation}
	\label{eq:NE}
	\begin{aligned}
		\mathbf{\overline{w}}(k) &=
		\begin{bmatrix}
			\overline{w}_1 \\ \overline{w}_2 \\ \overline{w}_3 	\end{bmatrix} (k) = \mathbf{L} \mathbf{e}_m(k) = 
		\begin{bmatrix}
			l_1 \\ l_2 \\ l_3
		\end{bmatrix}
		\mathbf{e}_m(k), \\
		\mathbf{e}_m(k) &= \mathbf{y}(k) - \mathbf{y}_m(k).
	\end{aligned}
\end{equation}

In~\eqref{eq:NE}, $\mathbf{y}_m$ is the estimated output of the internal model, $\mathbf{y}$ is the measurement output of the plant, while $\mathbf{L}$ collects the closed-loop predictor gains $l_k\mbox{, }k\,{=}\,1,\dots,3$, for the specific DC motor case. As a result, the closed-loop predictor model can be determined by designing and tuning the $l_k$ gains trading-off between the closed-loop stability and the estimation performance.

A static-feedback is selected for the design of noise estimator, and its stability analysis is already studied in \cite{Nanu2022}, Appendix B.

	\subsection{Asynchronous sampling time stability remarks}

From the experimental design perspective, the discrete-time eigenvalues of the control unit naturally change at each time-step. This is directly related to the nature of the NCS environment, characterised by highly variable sampling times. As a result, the adopted design solution was based on the design of a set of continuous-time eigenvalues first, focusing on expected stability and performance level. In particular, $ \mu_{_K} $ are the control law continuous-time eigenvalues, $ \mu_{_R} $ the reference dynamics eigenvalues and finally $ \mu_{_N} $ the predictor eigenvalues. Indeed, such a continuous-time set of eigenvalues, $\mu_i,\,i\,{=}\,1, \dots, n$, do not depend on the variable sampling time and can be then converted, at every time-step, into discrete-time eigenvalues $\lambda_i = e^{\mu_i T_{s,k}},\,i\,{=}\,1, \dots, n$, being $T_{s,k}$ the generic sampling time of the $k^{th}$ time-step.

\begin{figure}[thpb]
	\centering	\includegraphics[width=0.8\linewidth]{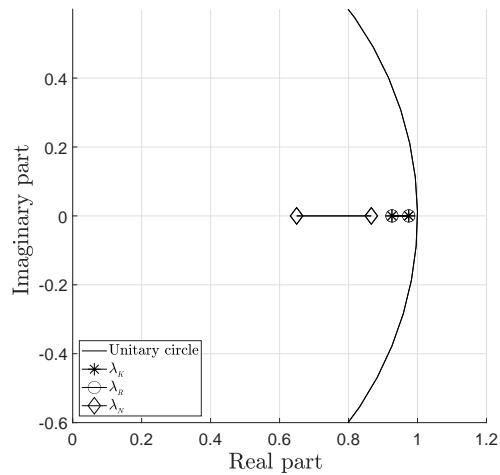}
	\caption[]{Experimental results: polar plot of the stability conditions of the control system, for $ T_s = [0.01, 0.03] \ s $}
	\label{fig:stab_polar_plot}
\end{figure}

In Figure \ref{fig:stab_polar_plot} the discrete-time domain eigenvalues are presented, for reference dynamics $ \lambda_{_R} = e^{\mu_{_R} T_s} $, control law $ \lambda_{_K} = e^{\mu_{_K} T_s} $ and predictor $ \lambda_{_N} = e^{\mu_{_N} T_s} $: it can be seen that, for the defined variation of $ T_s $, spanning from $0.01 \si{\second}$ to $0.03 \si{\second}$, the values are always inside the unitary circle, thus guaranteeing asymptotic stability.

\section{Additional experimental results}

In this section experimental tests are presented, which give additional results for the work in \cite{Nanu2022}.

	\subsection{EMC mobile robot test - NCS disturbance rejection effectiveness}

\begin{figure}[thpb]
	\includegraphics[width=\linewidth]{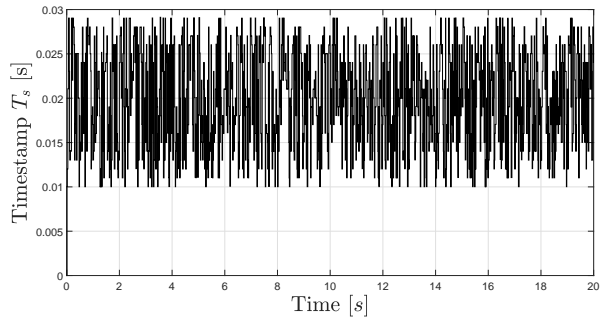}
	\caption{Experimental results: asynchronous timestamp with range $ T_s = [0.01, 0.03] \ \si{\second} $}
	\label{fig:right_DCmotor_timestamp}
\end{figure}

\begin{figure}[thpb]
	\includegraphics[width=\linewidth]{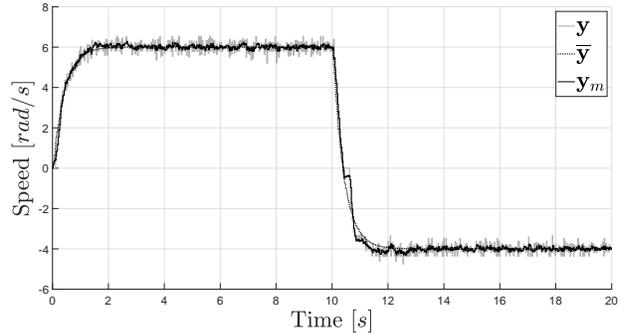}
	\caption{Experimental results: DC motor output speed estimation ($\mathbf{y}_m$) and comparison with reference ($\mathbf{\overline{y}}$) and measured ($\mathbf{y}$) values  - Disturbance rejection case}
	\label{fig:right_DCmotor_distrej_outputs}
\end{figure}

\begin{figure}[thpb]
	\centering
	\includegraphics[width=\linewidth]{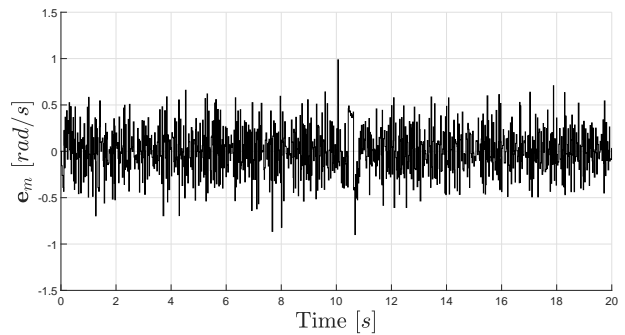}
	\caption{Experimental results: Tracking error $ \mathbf{e}_m $}
	\label{fig:RM_measured_em_bn}
\end{figure}

Figure \ref{fig:right_DCmotor_distrej_outputs} presents an experimental test with the right motor of the differential-drive robot (the results are analogous for the left motor), in the case of EMC disturbance rejection active. The motor output speed must follow a reference trajectory with the values of $ \SI{6}{\radian\per\second} $ and $ -\SI{4}{\radian\per\second} $, with a first-order dynamics in two different time periods. The sampling time is asynchronous and defined in the range $ T_s = [0.01, 0.03] \ \si{\second} $, as it can be seen in Figure \ref{fig:right_DCmotor_timestamp}. The measured $ \mathbf{y} $ and the estimated $ \mathbf{y}_m $ outputs are compared to understand the internal model observer estimation capabilities, quantified by the model error $ \mathbf{e}_m = \mathbf{y} - \mathbf{y}_m $, which is shown in Figure \ref{fig:RM_measured_em_bn}. The model error results to be in the range of the DC motor encoder maximum angular speed resolution error, $ e_{enc} = 0.8726 \ \si[per-mode=symbol]{\radian\per\second} $. This implies a remarkable and effective estimation performances, as it can be visually appreciated in Figure \ref{fig:right_DCmotor_distrej_outputs}, where $ \mathbf{y}, \mathbf{y}_m $ basically overlap.

	\subsection{EMC mobile robot test - Critical asynchronous sampling times}

\begin{figure}[thpb]
	\centering
	\includegraphics[width=\linewidth]{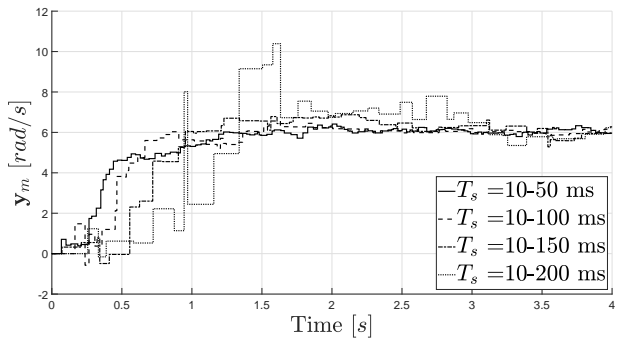}
	\caption[]{Experimental results: $ \mathbf{y}_m $ for several variable time-stamps $ T_s $}
	\label{fig:right_DCmotor_crit_cond}
\end{figure}

Figure \ref{fig:right_DCmotor_crit_cond} shows the estimated output of the EMC internal model $ \mathbf{y}_m $, for several asynchronous sampling times $ T_s $; spanning from a fixed minimum value of $ 10 $ ms, to a variable maximum value of $ [50,100,150,200] $ ms. For this experimental test, the reference motor speed is imposed to a fixed value of $ \mathbf{\overline{y}} = 6 $ rad/sec. The estimation of $ \mathbf{y}_m $ is effective and characterized by a negligible model error $ \mathbf{e}_m $, until a maximum value $ T_s = 150 $ ms, which is sufficiently high to cover a wide range of NCS delay and package loss variations. Only for higher sampling times, $ T_s > 150 $ ms, the model output $ \mathbf{y}_m $ starts to show a degradation in estimation performance.

	\subsection{EMC mobile robot test - EMC and PI comparison}

\begin{figure}[thpb]
	\centering	\includegraphics[width=\linewidth]{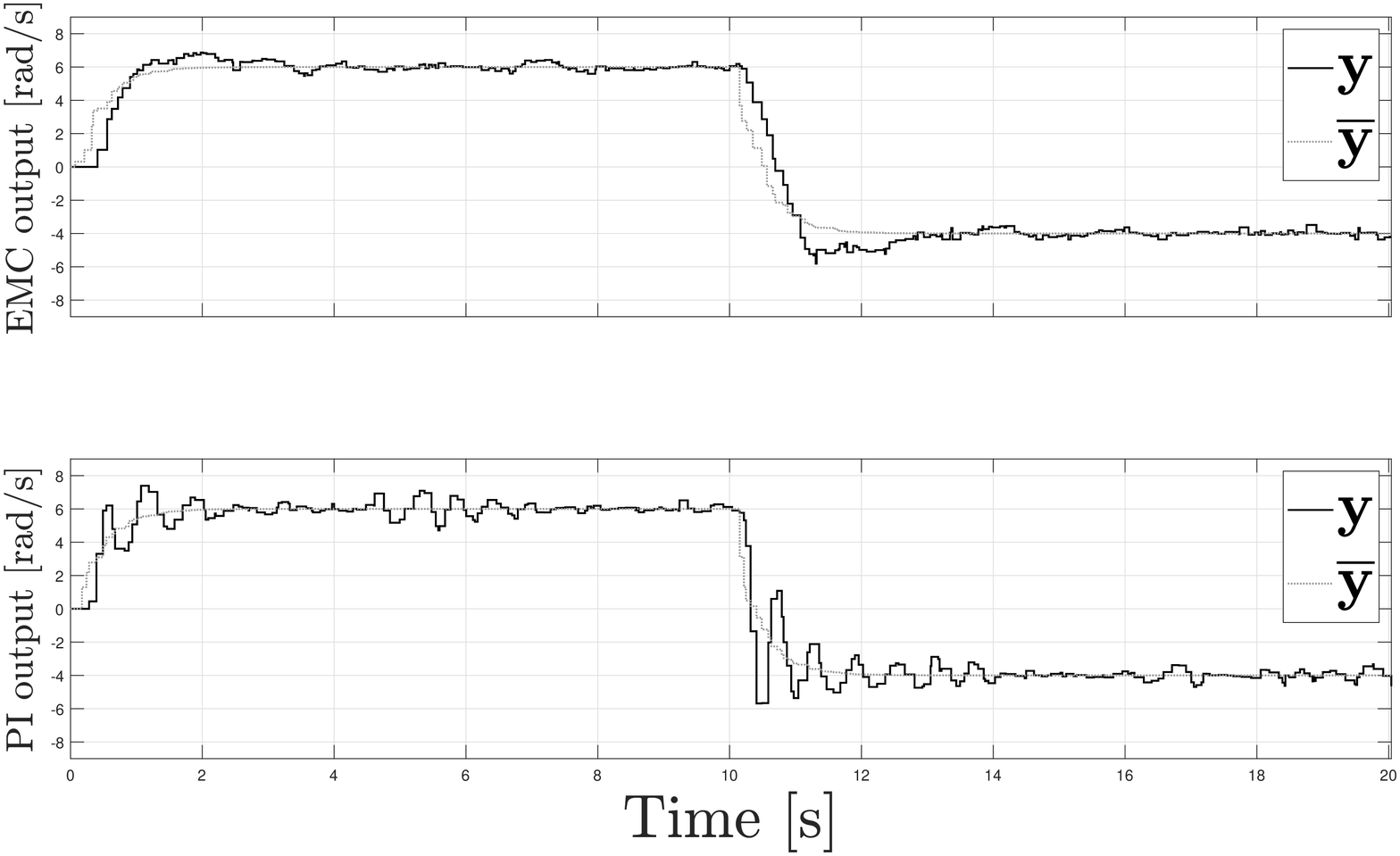}
	\caption[]{EMC and PI output speed comparison - Experimental test with timestamp $ T_s = [10-150] $ ms}
	\label{fig:EMC_vs_PI_outputs_10_150ms}
\end{figure}

\begin{figure}[thpb]
	\centering
	\includegraphics[width=\linewidth]{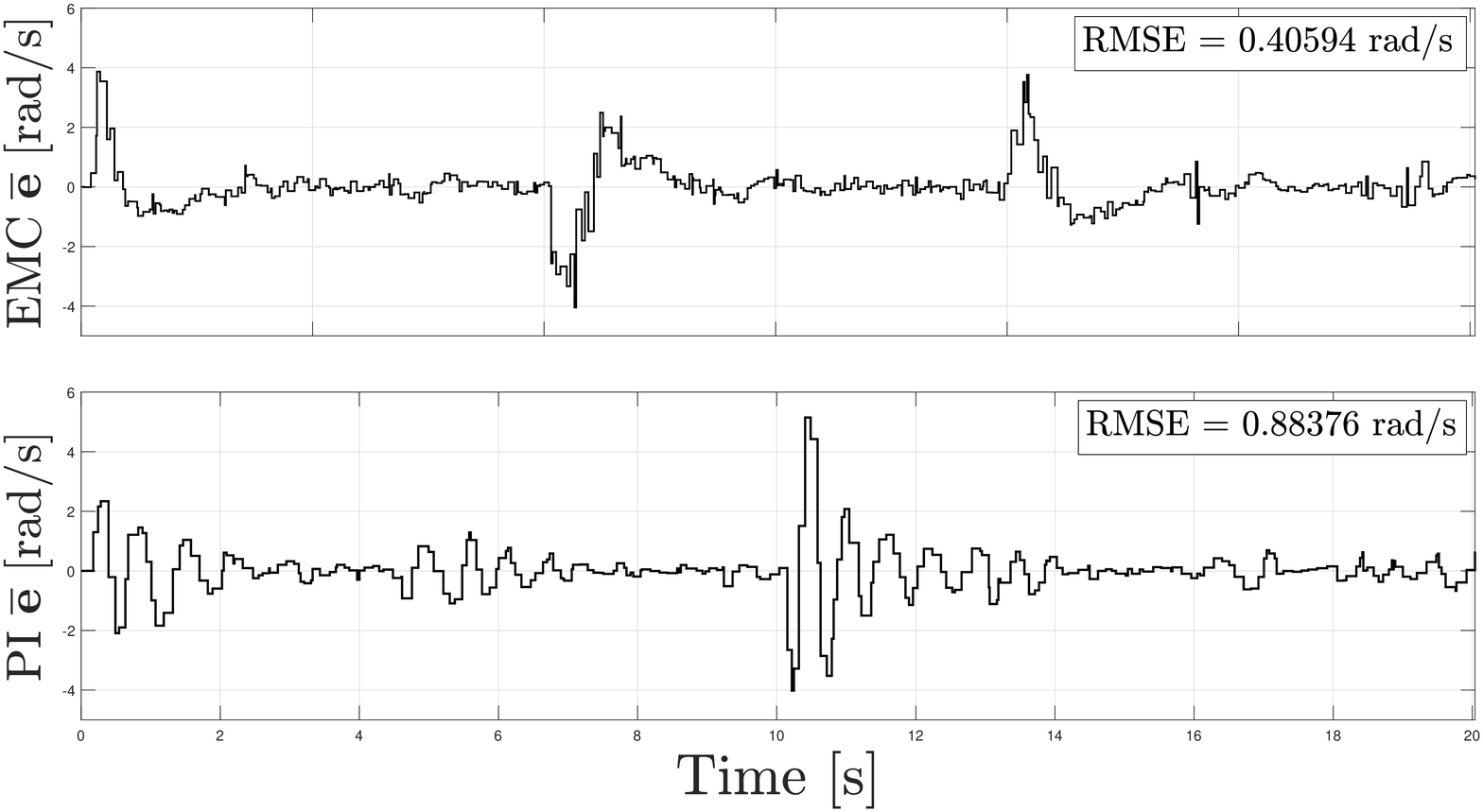}
	\caption{EMC and PI comparison: tracking error $ \mathbf{\overline{e}} $ with RMSE - Experimental tests with timestamp $ T_s = [10-150] $ ms}
	\label{fig:track_err_EMC_PI}
\end{figure}

\begin{figure}[thpb]
	\centering
	\includegraphics[width=\linewidth]{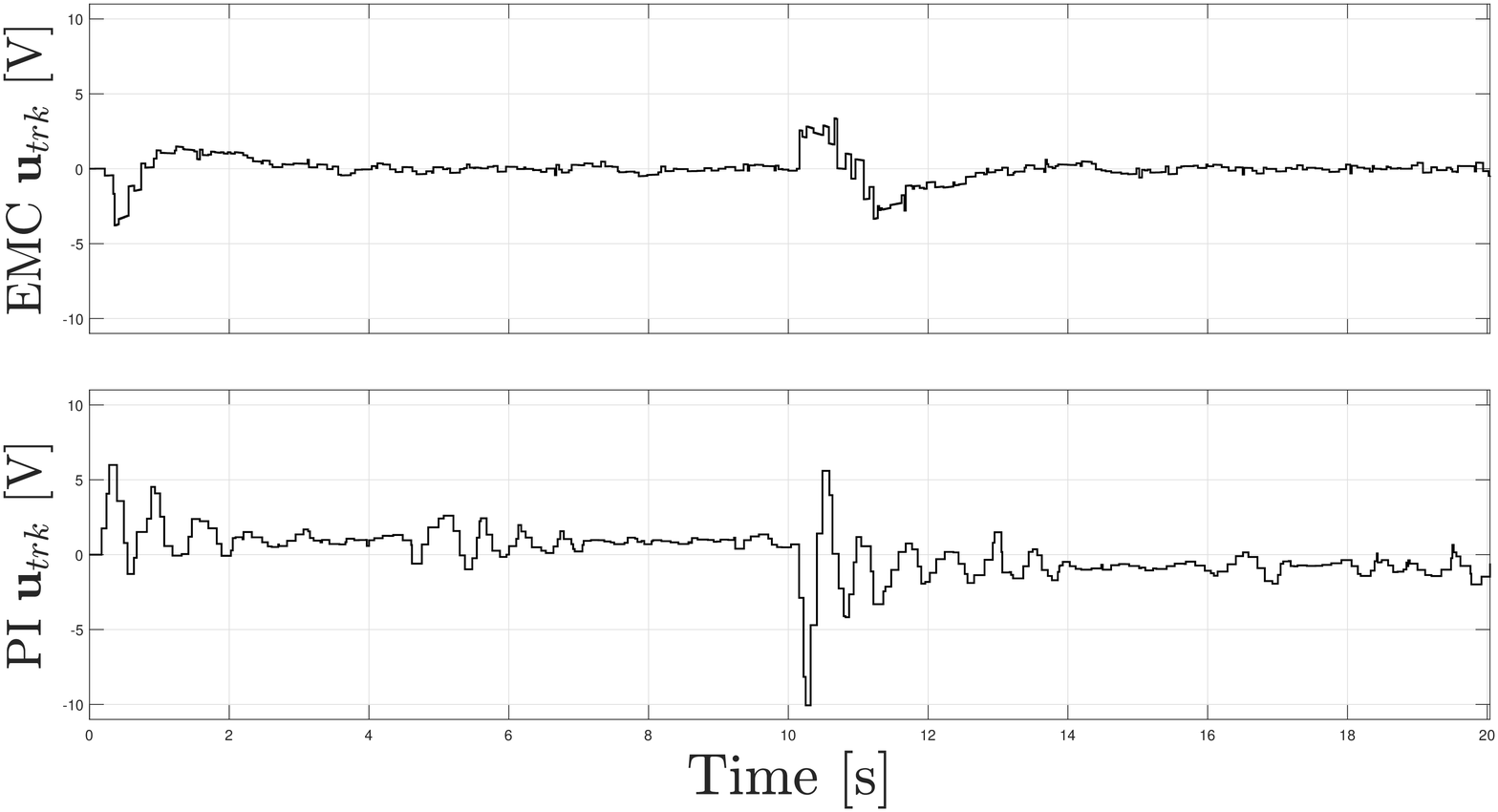}
	\caption{EMC and PI comparison: tracking control input $ \mathbf{u}_{trk} $ - Experimental tests with timestamp $ T_s = [10-150] $ ms}
	\label{fig:control_inputs_EMC_PI_10_150ms}
\end{figure}

In \cite{Nanu2022} a comparison between the EMC and a complete PID controller is made. However other experimental tests are conducted by considering a partial Proportional-Integral (PI) controller. Such a benchmark is adopted because, in a set of experimental tests, it is found that the absence of the derivative term led to more satisfactory control behaviours. The experimental comparison between the EMC and PI controllers, in critical varying time conditions, is shown in Figures from \ref{fig:EMC_vs_PI_outputs_10_150ms} to \ref{fig:control_inputs_EMC_PI_10_150ms}. The adopted variable timestamp is in the range of $ T_s = [10-150] $ ms, which introduces high delays in the control system and simulates an NCS scenario. The EMC and PI controller parameters are listed in Table \ref{tab:PI_EMC_comparison_par}.

\begin{table*}[!t]
	\centering
	\renewcommand{\arraystretch}{1.3}		
	% \extrarowheight
	\caption{EMC and PI controllers experimental test parameters}
	\label{tab:PI_EMC_comparison_par}
	\centering
	\begin{tabular}{lc}
		\hline
		Parameter & Value \\
		\hline
		Experimental test time step $ T_s \ [\si{\second}]$ & $[0.01, 0.15]$ \\
		\hline
		EMC Ref. dynamics eigenvalues $\mu_{_R}$ & $ \left[ -2.5647 \right] $ \\
		EMC Feedback eigenvalues $\mu_{_K}$ & $ \left[ -2.5647, -2.5647 \right] $ \\
		EMC Predictor eigenvalues $\mu_{_N}$ & $ \left[ -14.3842, -14.3842, -14.3839 \right] $ \\
		\hline
		PI proportional gain $ K_P $ & 1.35 \\
		PI integral gain $ K_I $ & 11.25 \\
		\hline
	\end{tabular}
\end{table*}

In Figure \ref{fig:EMC_vs_PI_outputs_10_150ms} the EMC control unit and a Proportional-Integral controller were compared in the mentioned experimental scenario. The measured output is $ \mathbf{y} $ and the reference to be tracked is $ \mathbf{\overline{y}} $. The PI controller results to have difficulties to track the reference output speed, even in steady-state conditions. Instead the EMC, after an initial transient with acceptable delays and overshoots, is able to follow $ \mathbf{\overline{y}} $ with small errors.

Such a behaviour is validated by Figure \ref{fig:track_err_EMC_PI}, where the Root Mean Square Error (RMSE) between $ \mathbf{\overline{y}} $ and $ \mathbf{y} $ is computed. In the EMC case, the error is $ RMSE_{EMC} = 0.40594 \ \si{\radian\per\second} $: significantly lower than the one for the PI controller, $ RMSE_{PI} = 0.88376 \ \si{\radian\per\second} $.

Finally in Figure \ref{fig:control_inputs_EMC_PI_10_150ms} the tracking control inputs for both the controllers were studied. In the EMC the control input $ \mathbf{u}_{trk} $ is lower in magnitude compared to the PI control input, and presents a more stable behaviour in steady-state conditions.

This experimental behaviour is driven by the structure of the EMC control block. Indeed, the EMC control input has 2 key parts: the feedback $ \mathbf{u}_{trk} $, leveraging the tracking error, and the disturbance rejection signal $ \mathbf{u}_d $. The last term, $ \mathbf{u}_d $, is driven by the disturbance dynamics estimated by the state observer, and counteracts all the disturbances affecting the plant behaviour; up to a certain bandwidth (set by the observer eigenvalues) \cite{Canuto2007}. This contributes in discharging the tracking control input $ \mathbf{u}_{trk} $, i.e. the feedback part of the control law, whose main tasks becomes counteracting residual effects not estimated by the observer \cite{novara2016control}. Conversely, in the PI unit the control is only managed by $ \mathbf{u}_{trk} $, hence leading to the unexpected control oscillations.

\section{Conclusion}
This work is intended to give additional studies and results to the main work \textit{Embedded Model Control of Networked Control Systems: an Experimental Case-study}, \cite{Nanu2022}. In this main work, the Embedded Model Control (EMC) methodology is applied to design a complete digital control unit for a differential-drive mobile robot controlled in a network controlled system (NCS), thus operating in a scenario characterised by a varying sampling time and asynchronous command execution.

At the beginning, this complementary paper presents a study on the stability analysis of the EMC, in the case of the differential-drive robot setup; namely the setup explored in \cite{Nanu2022}. Furthermore, experimental tests are added to those presented in \cite{Nanu2022}. The first test verified the EMC asynchronous timing disturbance rejection capabilities, in the case of the differential-drive robot setup. The second experiment aimed to test the EMC in critical asynchronous NCS scenarios, with high delays of the timestamp. The EMC resulted to have satisfactory control behaviours until high ranges of variable timestamp, thus validating the high practical applicability of the EMC control. Another experiment benchmarked the proposed EMC asynchronous architecture with a Proportional-Integral controller (a design showing a more satisfactory control behaviour than other control architectures, e.g. PID). The experimental benchmarking study showed that the EMC has a better tracking performance than the PI controller, thanks to the presence of the new disturbance rejection control term.

\bibliographystyle{elsarticle-num.bst}
\bibliography{library}

\end{document}